\documentclass[conference]{IEEEtran}
\input epsf
\usepackage{graphicx}
\usepackage{amsmath}
\begin{document}
\title{\LARGE Modeling of a tunable-barrier non-adiabatic electron pump\\
beyond the decay cascade model}
\author{\authorblockN{Vyacheslavs Kashcheyevs and Janis Timoshenko}
\authorblockA{Faculty of Physics and Mathematics, University of Latvia, LV-1586, Riga, Latvia\\E-mail: slava@latnet.lv}
}
\maketitle
\begin{abstract}
We generalize the decay cascade model of charge capture statistics for a tunable-barrier non-adiabatic electron pump
dominated by the backtunneling error at the quantum dot decoupling stage.
The energy scales controlling the competition between the thermal and the dynamical mechanisms for accurate trapped charge  quantization  are discussed.
Empirical fitting formula incorporating quantum dot re-population errors due to particle-hole fluctuations in the source
lead is suggested and tested against
an exactly solvable rate equation model.
\end{abstract}
\IEEEoverridecommandlockouts

\IEEEpeerreviewmaketitle

\section{Introduction}
Charge pumps based on tunable-barrier semiconductor quantum dots \cite{blumenthal2007a} can implement frequency-locked single-electron transfer with remarkable accuracy \cite{Giblin2012}, serving as potential building blocks for a quantum current standard. In a single-parameter non-adiabatic quantized pumping scheme \cite{Kaestner2007c,fujiwara2008}, the bottleneck for accurate current quantization is the initialization phase of the pumping cycle~\cite{Giblin2012,Kashcheyevs2010,Fricke2012}, in which the entrance barrier separating the quantum dot from the source lead is closed, trapping a certain number of electrons  electrons $n$ on the dot. These electrons are later ejected with high fidelity into the drain lead, producing a directed current $I=q_e f \langle{n}\rangle$, where $q_e$ is electron charge, $f$ is the pumping frequency (up to the gigahertz range \cite{blumenthal2007a,Giblin2012}) and $\langle{n}\rangle$ is the statistical average of the number of trapped electrons.

Decay cascade model~\cite{Kashcheyevs2010} has been successfully employed to describe the $I$-$V$ characteristics of a non-adiabatic quantized charge pump
using a fitting formula
\begin{equation}
    \langle{n}\rangle =\exp \bigl ( -e^{-\alpha(V-V_0)} \bigr) +\exp \bigl(-e^{-\alpha(V-V_0)+\delta_2} \bigr ) , \label{eq:CascadeClassic}
\end{equation}
where $V$ is the DC voltage on the exit barrier gate, and $\alpha$, $V_0$, and $\delta_2$ are the fitting parameters.
(We focus on the interplay of the first two charge states which limits $n$ to $0$, $1$ or $2$.)
The value of $\delta_2$ can be obtained from measurements of relatively low precision, and serves as a convenient
figure-of-merit  for predicting the ultimate flatness of the current quantization plateaux for a particular device.

There are certain limitations of the decay cascade model which this paper aims to address.
Experimentally, deviations of the current quantization steps from the double-exponential shape \eqref{eq:CascadeClassic} have been reported in high-precision measurements \cite{Giblin2012,Giblin2010a}. 
The original formulation of model~\cite{Kashcheyevs2010} considers the limit of an extremely rapidly rising energy of the quantum dot,
such that  $\langle n \rangle$ is set by the backtunneling dynamics only, and  $\delta_2$ is determined by the ratio of the decay rates $\Gamma_n$ for the first two charge states,
$\delta_2 \to \ln (\Gamma_2/\Gamma_1)$. In a later publication \cite{Fricke2012}, a more general model of the non-equilibrium charge capture has been
put forward which includes also the addition energy $E_c=\mu_2-\mu_1$ and the temperature $T$. It has been shown
that  Eq.~\eqref{eq:CascadeClassic} is reproduced by the general model in the limit of $T\to 0$.
However, finite-temperature corrections and the physical parameters controlling them have not been characterized.

In this paper, we show that the empirical figure-of-merit $\delta_2$ of the decay cascade model includes both temporal and energetic separation between the electrons escaping back to the source:
\begin{equation} \label{eq:Delta2}
  \delta_2 = \ln \frac{\Gamma_2}{\Gamma_1} + \frac{E_c}{\Delta_{\text{ptb}}} .
\end{equation}
Here $\Delta_{\text{ptb}}$ is the plunger-to-barrier ratio \cite{VKJT2012} of the entrance barrier gate.
$\Delta_{\text{ptb}}$ is equal to shift of the energy levels of the quantum dot during the time  $\tau$ it takes for the tunneling rate to be reduced $e=2.718\ldots$ times; it is a measure of cross-talk in the design of the energetic (``plunger'') and the barrier functions of the gate.

At finite temperatures we find that the corrections to Eq.~\eqref{eq:CascadeClassic} are controlled by the ratio $k T/ \Delta_{\text{ptb}}$ ($k$ is the Boltzmann constant).
Decay cascade formula  \eqref{eq:CascadeClassic} remains a good approximation for  $k T/ \Delta_{\text{ptb}} < 0.1$. At higher temperatures the following
\emph{ansatz} for the current quantization plateaux (defined by the range  $0.9<\langle n \rangle <1.1$) describes well the crossover between the backtunneling and
the thermal error mechanisms:
\begin{equation} \label{eq:tailsnatz}
 \langle n \rangle  = 1 -e^{-\alpha_{1} (V-V_{1})} + e^{+\alpha_{2} (V-V_2)} \, .
\end{equation}
The ratio $\alpha_{1} / \alpha_2$ extracted from numerical data is close to $k T/ \Delta_{\text{ptb}}$,
as expected from our recent analysis of fluctuations in a model of a single quantum level~\cite{VKJT2012}.

\section{Model}
We employ the sequential tunneling master equation with time-dependent rates \cite{Fricke2012},
\begin{eqnarray}
  dP_2^{}/dt & =& - W_2^{-} P_2^{} +W_1^{+} P_1^{} , \label{eq:master-1}\\
  dP_1^{}/dt & =&  + W_2^{-} P_2^{} - (W_1^{+}+W_1^{-}) P_1^{} + W_0^{+} P_0^{}, \label{eq:master-2}
\end{eqnarray}
where $W^{\pm}_n$ describes addition ($+$) or removal ($-$) of a single electron to a charge state $n$, and $P_n(t)$ is
the non-stationary probability distribution subject to normalization $P_0+P_1+P_2=1$. Transition rates are subject to a (generalized)
detailed balance condition, $W^{-}_n/W^{+}_{n\!-\!1}=e^{-(\mu_n-\mu)/kT}$, which defines the corresponding
time-dependent chemical potentials $\mu_n(t)$ of the charge state $n$  with respect to the grounded potential of the source lead, $\mu=0$.
The total rate $\Gamma_n=W^{-}_n+W^{+}_{n\!-\!1}$ is dominated by the exponential gate-voltage (and hence, time) dependence
of the tunneling matrix elements due to the field effect in the semiconductor constituting the closing entrance barrier between the source and the quantum dot.
Following \cite{Kashcheyevs2010,Fricke2012,VKJT2012} we postulate
\begin{eqnarray}
  \mu_1(t)    & = & \varepsilon_c(V) +  \Delta_{\text{ptb}} t/\tau , \label{eq:mu}\\
  \Gamma_1(t) & = & e^{-t/\tau}/\tau , \label{eq:Gamma}
\end{eqnarray}
around the first charge state decoupling time $t=0$.
The second state is separated from  the first by constant and positive $\Gamma_2/\Gamma_1$ and $E_c$.
The charge capture counting statistics \cite{Fricke2012} is given by the large-time limit of $P_n(t)$
starting from a grand canonical equilibrium at $t=t_0$. For $t_0 \ll -\tau$ the initial condition gets forgotten due to
adiabaticity of stochastic dynamics \eqref{eq:master-1}-\eqref{eq:master-2} with $\Gamma_n(t) \gg 1/\tau$ \cite{VKJT2012}.

Similar to the time-dependent entrance gate voltage which creates the time-dependencies \eqref{eq:mu} and \eqref{eq:Gamma},
a DC voltage $V$ on a tuning (e.g., exit) gate is expected to affect energies linearly and tunneling rates exponentially.
Both effects result in a linear shift, $\varepsilon_c(V)/\Delta_{\text{ptb}}=-\alpha (V-V_0) $ \cite{Kashcheyevs2010,VKJT2012}.

\section{Results}
\begin{figure}
\begin{center}
\includegraphics[width=0.8\columnwidth]{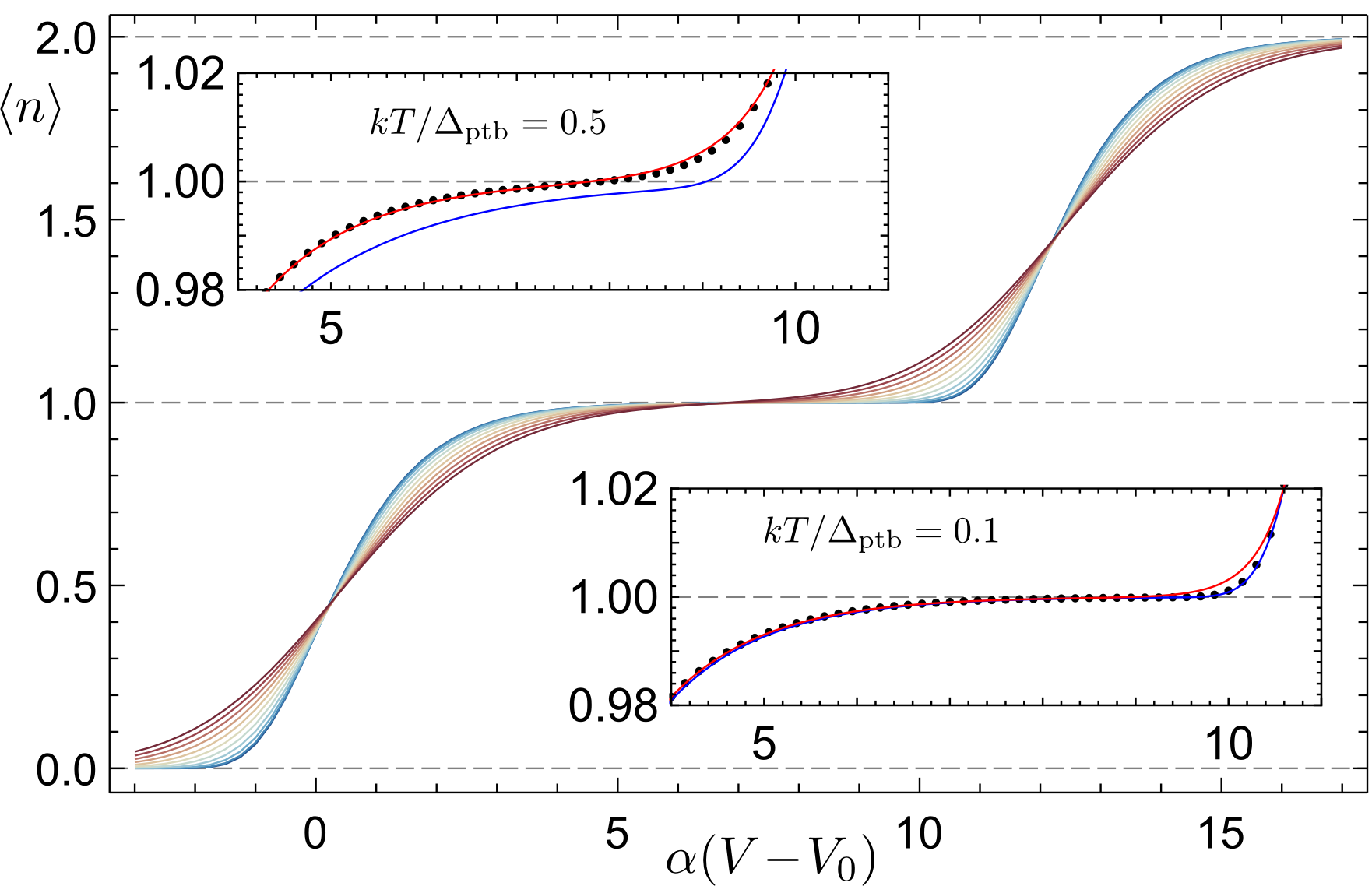}
\end{center}
  \caption{Average trapped charge $\langle n \rangle$ as function of $V$ for a uniform set of temperatures  $kT/\Delta_{\text{ptb}}=0$ to $1$ (blue to red, decreasing steepness).   Cascade parameters $\delta_2 =12$ and $E_c= 6 \Delta_{\text{ptb}}$, voltage rescaling parameters fixed to $V_0=0$ and $\alpha=1$.
  Insets show examples of fitting the pure decay cascade formula \eqref{eq:CascadeClassic} (blue, lower curve) and the fluctuation \emph{anstaz} \eqref{eq:tailsnatz} (red, upper curve) to the numerical solution (dots).\label{fig:Tup}}
\end{figure}

\begin{figure}
\begin{center}
\includegraphics[width=0.7\columnwidth]{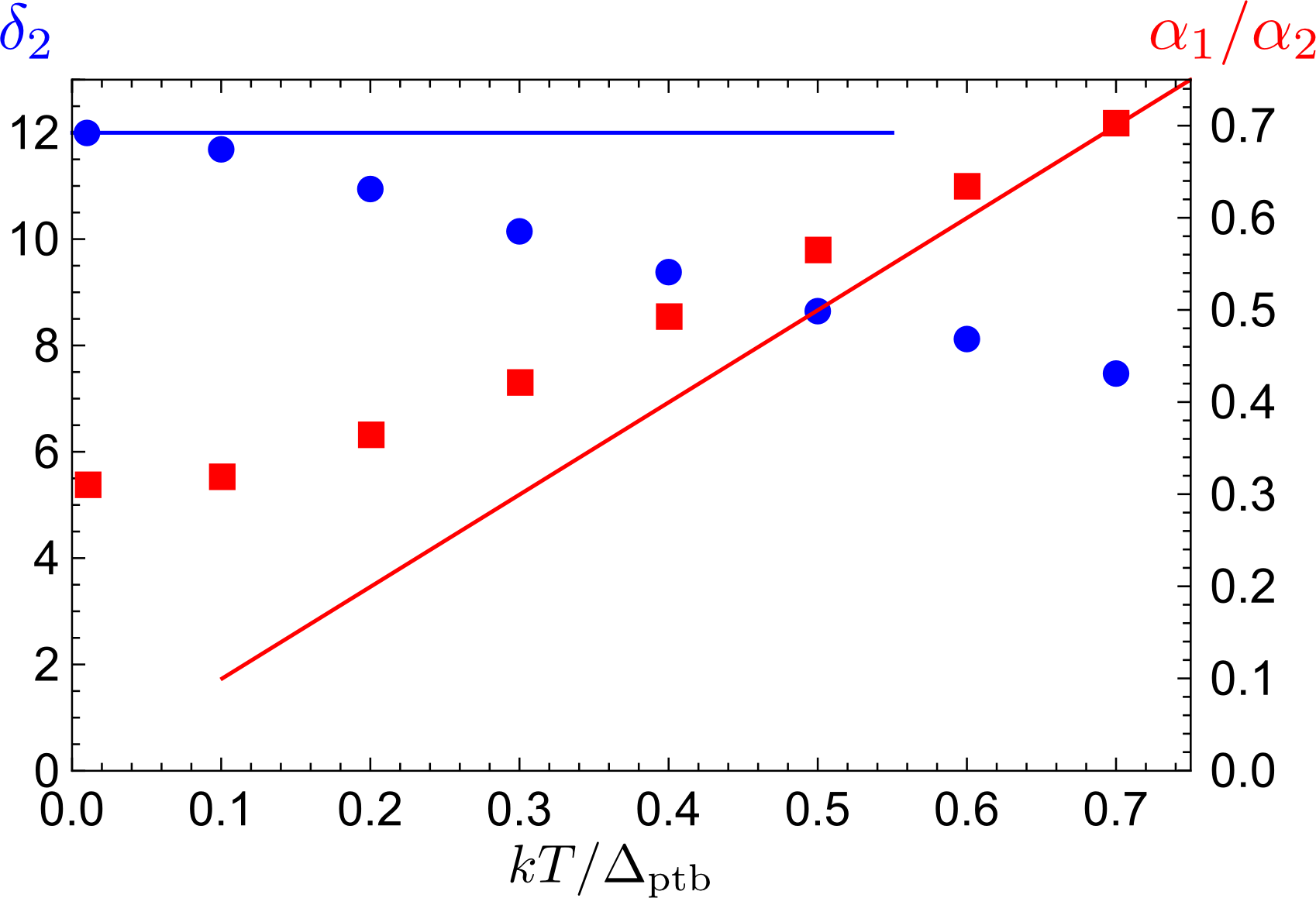}
\end{center}
  \caption{Results of fitting the decay cascade model (blue circles, left axis) and the fluctuation \emph{ansatz} (red squares, right axis) to the exact numerical solution as a function of temperature $kT/\Delta_{\text{ptb}}$. Model parameters same as in Fig.~\ref{fig:Tup}. Straight lines guide the eye to $\delta_2 = 12$ and $\alpha_1/\alpha_2 =kT/\Delta_{\text{ptb}}$.\label{fig:fit}}
\end{figure}

The zero-temperature solution reported in \cite{Fricke2012} in terms of the parametrization \eqref{eq:mu} and \eqref{eq:Gamma} results
in $\langle n \rangle =\sum_n n P_n$ given by Eq.~\eqref{eq:CascadeClassic} with $\delta_2$ given by Eq.~\eqref{eq:Delta2}.
Increasing $kT$ brings about gradual rounding of the steps and increase of the slope at the flattest part of the plateaux, see Fig.~\ref{fig:Tup}.
The decay cascade model formula \eqref{eq:CascadeClassic} gives a better fit than the fluctuation \emph{ansatz} \eqref{eq:tailsnatz} up to $k T \approx 0.1 \Delta_{\text{ptb}}$ (see the lower inset in Fig.~\ref{fig:Tup}), while at higher temperatures Eq.~\eqref{eq:tailsnatz} rather abruptly becomes a much better approximation  (see the upper inset in Fig.~\ref{fig:Tup}). When the fits are good, the values given by Eq.~\eqref{eq:Delta2} for $\delta_2$
and  $\alpha_{1} / \alpha_2 \approx kT/\Delta_{\text{ptb}}$ are recovered for the models \eqref{eq:Delta2} and \eqref{eq:tailsnatz},  respectively, as demonstrated in Fig.~\ref{fig:fit}.

\section{Conclusion}
Decay cascade model predictions are robust against thermal fluctuations up to an energy threshold which can be quantified in terms of a plunger-to-barrier ratio
intrinsic to the non-adiabatic single-gate charge pump design. Fits of the current quantization plateaux to a simple sum of two exponentials may help in investigation of the physical error mechanisms, and contribute to improved metrological performance of a quantum current standard.

\section*{Acknowledgment}
This work is supported by Researcher Excellence Grant SIMB07-REG4
within the European Metrology Research Program (EMRP) and by Project no.\ $146/2012$ from the Latvian Science Council.
The EMRP is jointly funded by the EMRP participating countries within EURAMET and the European Union.


\end{document}